\documentclass[note,traditabstract]{aa}

\usepackage{graphicx}
\usepackage{natbib}
\usepackage{amssymb}
\bibpunct{(}{)}{;}{a}{}{,}

\begin{document}

\title{Searching for an EBL attenuation signature in the\\ Fermi/LAT 1st year catalog data}

\author{M. Raue\inst{1}}

\institute{
Universit\"at Hamburg, Institut f\"ur Experimentalphysik, Luruper Chaussee
149, D 22761 Hamburg, Germany
}

\offprints{M. Raue, \email{martin.raue@desy.de}}

\date{Received  / Accepted}

\abstract {Observations of distant sources of high-energy (HE) $\gamma$-rays are affected by attenuation resulting from the interaction of the $\gamma$-rays with low energy photons from the diffuse meta-galactic radiation fields at ultraviolet (UV) to infrared (IR)  wavelengths (Extragalactic Background Light; EBL). Recently, a large data-set of HE observations from the 1st year survey of the Large Area Telescope (LAT) instrument on-board of the Fermi satellite became available, covering an energy range from 100\,MeV up to 100\,GeV. In this paper, the potential of such large HE data-sets to probe the density of the EBL - especially in the UV to optical - is explored. The data from the catalog is investigated for an attenuation signature in the energy range 10-100\,GeV and the results are compared with the predictions from EBL model calculations. No clear signature is found. The statistics are still limited by (1) the sensitivity of Fermi/LAT to detect sources above 10\,GeV, (2) the number of firmly identified sources with known redshift, both which will improve over the coming years.
}



\maketitle

\section{Introduction}

Measuring the energy spectrum of distant sources of high-energy $\gamma$-rays (100\,MeV - 100\,GeV; HE) provides a unique method to probe the densities of extragalactic diffuse photon fields at ultraviolet to infrared wavelengths (extragalactic background light; EBL). The HE $\gamma$-rays interact with the low energy photons of the EBL via the pair-production process leaving an energy dependent attenuation signature in the spectra \citep{nikishov:1962a,gould:1967a}. The strength of this attenuation and its spectral behavior depends on the properties of the EBL (i.e. its  spectral number density).

Very-high energy (VHE; E$>$100\,GeV) $\gamma$-ray observations of distant sources performed with ground-based instruments have successfully been used to limit  the EBL density  \citep[e.g.][]{aharonian:2006:hess:ebl:nature,mazin:2007a,albert:2008:magic:3c279:science}. Unfortunately, typical ground-based VHE instruments have limited observation time ($\sim1000$\,h/year) and field of view ($2.5-5^\circ$ diameter) and therefore the number of known extragalactic VHE  sources is  small (O(30)). EBL studies using VHE data, therefore, often rely on the observations of a few sources (sometimes even a single source).

Different possibilities arise from observations with the Large Area Telescope (LAT; \citealt{atwood:2009:fermi:lat:technical}) on board the Fermi satellite. The detector has a large field of view (2.4\,sr) and is operated mostly in survey mode covering the full sky every 3\,h. Fermi/LAT observations cover the energy range from 100\,MeV to 100\,GeV and thereby enable to probe the EBL at ultraviolet to optical wavelengths.
For the redshift range investigated in this paper (up to $z=2-3$), the energy spectrum below 10\,GeV is essentially not affected by EBL attenuation.
%
Strong contributions from early stars to the EBL density could, in principal, lead to some attenuation in this energy range \citep{kashlinsky:2005b} but the claimed experimental evidence for such a high contribution (see e.g. \citealt{dwek:1998b,gorjian:2000a,cambresy:2001a,matsumoto:2005a} and \citealt{Kashlinsky2005:EBLReview} for a review) has been challenged on theoretical \citep[e.g.][]{madau:2005a} and experimental \citep[e.g.][]{dwek:2005c,aharonian:2006:hess:ebl:nature,thompson:2007a} grounds.
%
Therefore, it is possible to (i) sample parts of the intrinsic spectrum emitted at the source and (ii) to observe sources up to higher redshift compared to VHE observations.\footnote{Source with a redshift of $z\gtrsim0.5$ are expected to be severely attenuated due to the interaction with the EBL at energies above 100\,GeV.
}
With the recent release of the Fermi/LAT 1 year point source catalog (1FGL; \citealt{abdo:2010:fermi:1fgl}\footnote{http://fermi.gsfc.nasa.gov/ssc/data/access/lat/1yr\_catalog/}; CAT1 in the following) and the associated first LAT AGN catalog (1LAC; \citealt{abdo:2010:fermi:1stagncat}; CAT2 in the following) a large high-energy $\gamma$-ray data-set became available. The catalog contains $\sim$700 $\gamma$-ray sources associated with extragalactic objects and enables, for the first time, to perform a population study for the global EBL attenuation effect on a sufficiently large data-set \citep[e.g.][]{chen:2004a}.

In this paper the 1FGL data-set will be examined for a signature from EBL attenuation. The methods used for the study are discussed in Sec.~\ref{methods}, results are presented in Sec.~\ref{results} and Sec.~\ref{summary} gives a summary and conclusions.
For the calculations in the paper the following values for the cosmological parameters are adopted: $h=\Omega_\Lambda=0.7$ and $\Omega_M=0.3$.

\section{Methods and data sample}\label{methods}

\begin{figure}[tb]
\centering
\includegraphics[width=0.45\textwidth]{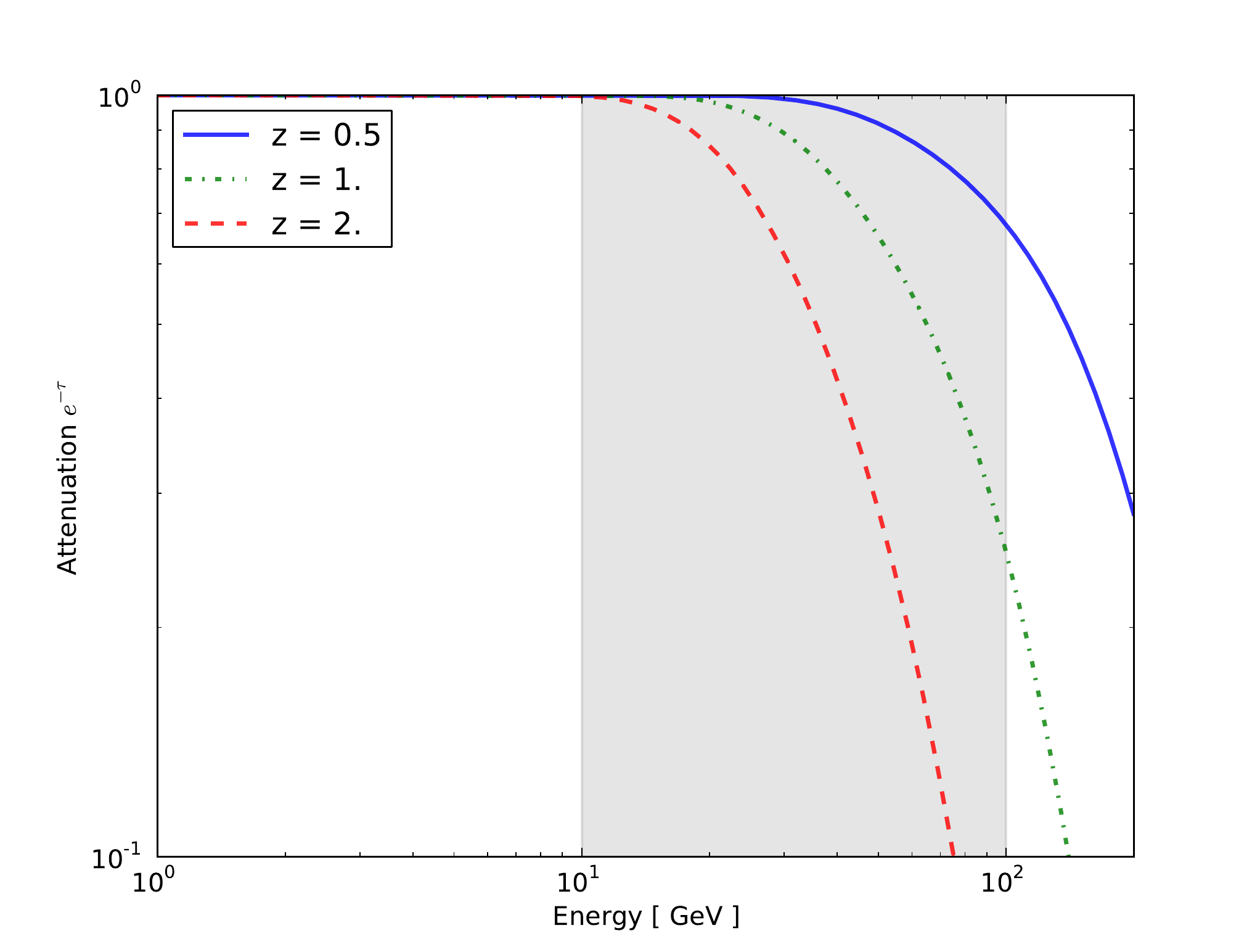}
\caption{Energy dependent attenuation resulting from the interaction of HE/VHE $\gamma$-rays with the low energy photons from the EBL for different redshifts $z$ (EBL model: KN low from  \citealt{kneiske:2002a}). 
The grey shaded area gives the energy range of the highest energy bin in the 1FGL catalog which will be investigate for attenuation features.
}
\label{Fig:EBLAttenuation}
\end{figure}

The 1FGL catalog lists for each source the integral flux for 5 bins in energy (0.1-0.3\,GeV, 0.3-1\,GeV, 1-3\,GeV, 3-10\,GeV, and 10-100\,GeV) and the results of the fit of a power law to the energy spectrum over the full energy range from 0.1 to 100\,GeV. Fig.~\ref{Fig:EBLAttenuation} displays the EBL attenuation for the low EBL model from \citet{kneiske:2002a} for the energy range 1\,GeV to 200\,GeV for several redshifts. It can be seen that significant EBL attenuation is only expected for the flux in highest energy bin. The sources in the catalog will therefore be examined for an attenuation signature in the highest energy bin (i.e. a deviation from the behavior at lower energies).
The main caveat for this type of approach is that intrinsic curvature towards the highest energies is expected for many emission models and has been detected in several sources for energies below 10 GeV. Therefore, no (strong) conclusion on the EBL attenuation can be drawn from the fact that the flux in the last bin is attenuated. On the other hand, if no attenuation is detected, this can be used to set limits on the attenuation strength and therefore the EBL density.
To search for a possible attenuation signature in the highest energy bin two different methods will be employed:

\begin{enumerate}
\item \textit{Power law fit.}
For this method, the power law fit to the energy spectrum is assumed to be a good representation of the overall spectrum emitted at the source. Due to the shape of the energy spectra\footnote{Most AGN energy spectra in the 1FGL are well described either by power laws with $\Gamma > 2$ or log parabola spectra.} and the sensitivity of the detector (best sensitivity in the 0.1-0.3\,GeV energy range) the spectral fit result for the full energy range are dominated by the photons with energies $< 10$\,GeV. The integral flux in the energy range 10 to 100\,GeV as calculated from the power law fit $F_{IntPL}$ can therefore be considered a good representation of the flux for the un-attenuated spectrum. The ratio of the measured flux in the last bin $F_{10-100\,GeV}$ and $F_{IntPL}$ is used as an indicator to quantify the strength of the attenuation. The main caveat for this approach is that it relies on the power law fit to be a good representation of the overall spectrum.
\item \textit{Flux stacking.}\, The flux in the highest energy bin $F_{10-100\,GeV}$ is compared to the integrated flux at energies $<$10\,GeV $F_{0.1-10\,GeV}$ by summing up the flux in the bins  0.1-0.3\,GeV, 0.3-1\,GeV, 1-3\,GeV, and 3-10\,GeV. The ratio between the two integral fluxes $F_{10-100\,GeV}$/$F_{0.1-10\,GeV}$ is used as indicator for an attenuation signature. This ratio depends on the spectral shape of the source and therefore, to be able to compare and combine the results for different sources, a sample of intrinsically similar sources is needed. While the BL Lac type sources show a wider spread of spectral shapes (spectral indices), the spectral shapes of the flat spectrum radio quasars (FSRQs) in the Fermi/LAT range are remarkably similar, displaying a narrow distribution of spectral indices with a mean of $\Gamma \sim 2.5$ (see CAT2 Fig.~12). Therefore, only spectra from FSRQ type sources will be used for this type of analysis.
\end{enumerate}

EBL attenuation is a redshift dependent process. A signature should therefore emerge when analyzing the evolution of the attenuation effect with redshift (for caveats see \citealt{reimer:2007a}). The 1FGL catalog does not contain redshifts, but these can be found (for the associated sources) in the 1LAC catalog. For the analysis the two catalogs are merged based on the catalog names using only the high quality data sample from the a 1LAC catalog (clean sample, see CAT2).
Of the 1451 sources in the combined catalog 287 ($\sim20$\%) have a significant detection in the last energy bin, 390 ($\sim27$\%) have a redshift associated, and 72 ($\sim5$\%) have both. 57 of these source do not show significant curvature (curvature index $<$11.34, see CAT1), leaving 15 sources with significant curvature in the spectrum. For the 2nd method, only FSRQ type sources will be used, which do account for 12 sources without and 6 sources with significant curvature. All numbers are  summarized in Tab.~\ref{Tab:DataSet}. The number of sources used in this study is only a small fraction of the total number of sources in the 1FGL. How this impacts the results and what to expect in the future will be discussed in Sec.~\ref{results} and Sec.~\ref{summary} respectively.

\begin{table}[tb]
\begin{center}
\begin{tabular}{lrr} \hline \hline
 & Number & Percentage \\ \hline
All & 1451 & 100\% \\ 
 $F_{10-100\,GeV}$ & 287 & 20\% \\
 z & 390 & 27\% \\
$F_{10-100\,GeV}$  + z & 72 & 5\% \\ 
$F_{10-100\,GeV}$  + z + NC & 57 & 4\% \\ 
$F_{10-100\,GeV}$  + z + C &  15 & 1\% \\ \hline
FSRQs & & \\
$F_{10-100\,GeV}$  + z + NC & 12 & $<$1\% \\ 
$F_{10-100\,GeV}$  + z + C &  6 & $<$1\% \\ \hline
\end{tabular}
\end{center}
\caption{Number of sources in the 1FGL catalog for specific criteria:  $F_{10-100\,GeV}$ - significant flux in the highest energy bin; z - known redshift; NC - no significant curvature detected in the spectrum; C - significant curvature detected in the spectrum; FSQR - source is of type FSRQ.}\label{Tab:DataSet}
\end{table}
 
\section{Results}\label{results}
  
\begin{figure}[tb]
\centering
\includegraphics[width=0.45\textwidth]{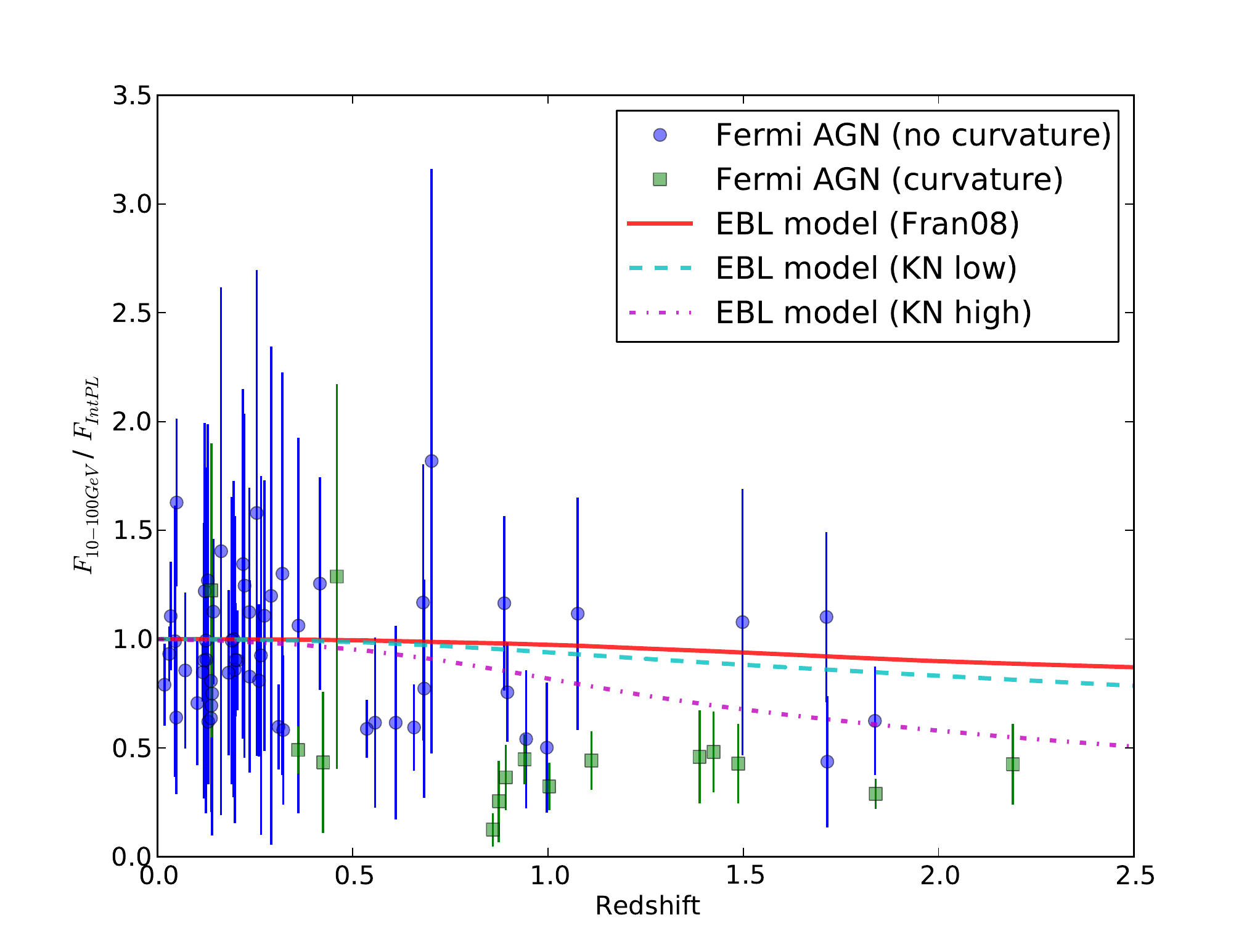}
\caption{Ratio of the integral flux in the highest energy bin $F_{10-100\,GeV}$ and the flux expected from the power law fit $F_{IntPL}$ versus redshift for AGN sources from the 1FGL catalog (method 1). For comparison the expectation for different EBL models is plotted. Error bars are 1$\sigma$ and include the uncertainty due the uncertainty in the power law fit.}
\label{Fig:Results1}
\end{figure}
 
\paragraph{Power-law fit.}
Figure~\ref{Fig:Results1} shows the ratio $F_{10-100\,GeV}$/$F_{IntPL}$ versus redshift for the source samples described in the previous section. For comparison, the expectation for the EBL attenuation for three different EBL models is also shown (\citealt{franceschini:2008a} = Fran08; KN low/high = low/high model from \citealt{kneiske:2002a}).
The models are chosen to cover some range in attenuation with the Fran08 model being at the low end, describing the EBL density derived from galaxy counts, and the KN high model resulting in the strongest attenuation but overproducing the results from galaxy counts.
The integrated attenuation for the EBL models is calculated under the assumption that the intrinsic spectrum follows a power law with spectral index $\Gamma = 2.5$.\footnote{The change in integrated attenuation when using $\Gamma = 2$ or $\Gamma = 3$ is less then 10\%.} For redshifts $<0.5$, where the majority of the sources in the sample is located (63\%), the expected EBL attenuation signature is less then 10\%. For redshift $z = 2.5$ up to 50\% attenuation is expected for $F_{10-100\,GeV}$. For the sources without intrinsic curvature the errors on the flux ratio are of the order of 0.2 to 0.5. The errors include the uncertainty in the parameters of the power law fit. At lower redshifts the flux ratios scatter around 1 as expected. At higher redshifts ($z>1.25$) only very few sources are available (4) and no clear attenuation signature is visible (i.e. sources following the trend of the theoretical predictions).
For completeness, the results for the sources with curved spectra are also shown. The curvature affects the power law fit and therefore the absolute value of the attenuation has to be taken with caution since it is likely underestimated. As expected, the sources with significant curvature show deviation from the power law fit for $F_{10-100\,GeV}$. For redshift $>0.5$ all curved sources show a concave behavior (attenuated last bin) and they all lie below the expectation for the EBL model with the highest attenuation used in this study
 
\begin{figure}[tb]
\centering
\includegraphics[width=0.45\textwidth]{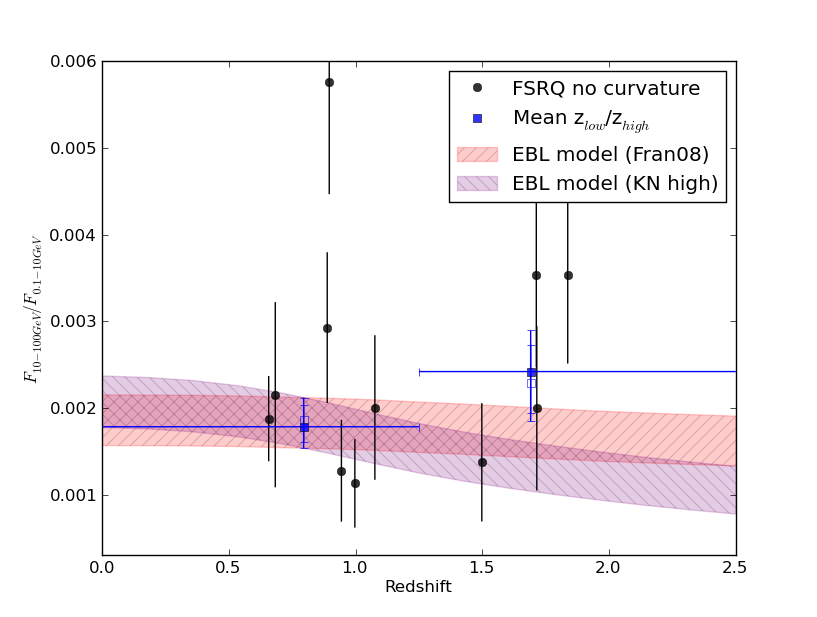}\\
\includegraphics[width=0.45\textwidth]{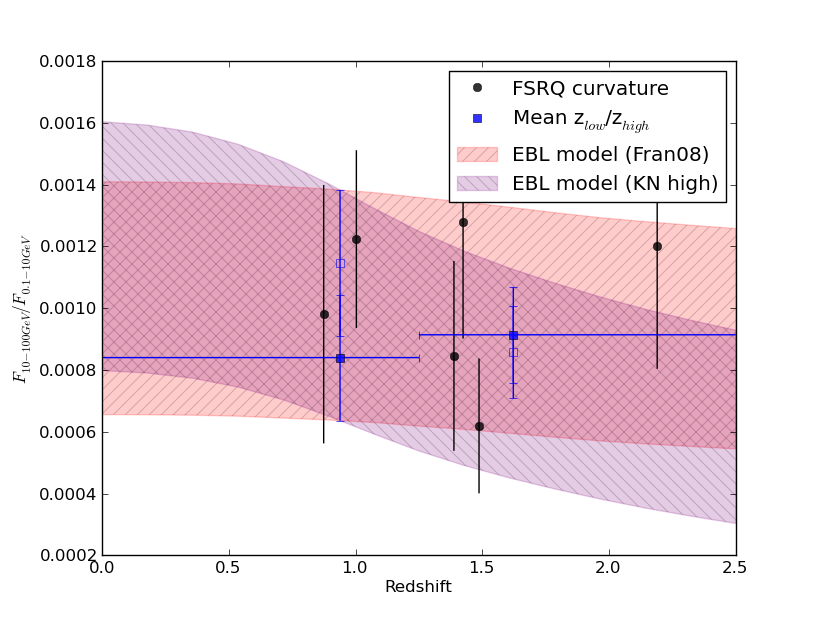}\caption{Ratio of the fluxes in the energy range 10-10\,GeV and 0.1 to 10\,GeV $F_{10-100\,GeV}$/$F_{0.1-10\,GeV}$ versus redshift for FSRQs from the 1FGL catalog (upper panel: without curvature; lower panel: with curvature). In blue the mean values for two bins in redshifts are shown ($z < 1.25$ and $z>1.25$; filled symbols: flux stacking; open symbols: average; see text for details). The colored bands denote the expectation for two different EBL models normalized to the low redshift mean value. In the upper panel one point at $z=0.22$ with $F_{10-100\,GeV}$/$F_{0.1-10\,GeV} = 0.014 \pm 0.007$ is not shown to improve the plots readability.}
\label{Fig:Results2}
\end{figure}

\paragraph{Flux stacking.} The $F_{10-100\,GeV}$/$F_{0.1-10\,GeV}$ values from the flux stacking for FSRQs without and with significant curvature are shown in Fig.~\ref{Fig:Results2} upper and lower panel respectively. Also shown are the mean values for two bins in redshift ($z < 1.25$ and $z > 1.25$). The mean is calculated using two different methods: (1) The $F_{10-100\,GeV}$/$F_{0.1-10\,GeV}$ values for different sources in one redshift bin are averaged, weighted with $1/\sigma^2$ (open blue squares). (2) For all sources in one redshift bin, first, the fluxes are averaged in the different energy bins (again weighted with $1/\sigma^2$) and then summed to get the total fluxes for the ranges 0.1-10 and 10-100\,GeV (filled blue squares). The second method has the advantaged that sources, which do not have a significant detection in one of the energy bins, do not distort the summed flux. The mean values are plotted at the mean redshift of the sources in the bin. For comparison, the expectations for different EBL models are shown. Since $F_{10-100\,GeV}$/$F_{0.1-10\,GeV}$ is a relative value the EBL model predictions are normalized to the mean value in the low redshift bin (at the mean redshift). Shown is the outer value of the one sigma error band for the two different mean values. As noted earlier, the total number of sources in the analysis is low, in particular for the case of the curved spectrum (6 sources in total). For the individual sources the spread is rather large and no clear trend is visible.

When considering the mean values for the two bins in redshift it can be noted that, for the source with no curvature (upper panel), the mean in the high redshift bin lies above the value for the low redshift bin. While this is not significant it might indicate a change in the intrinsic spectral properties with redshift. The mean value for the high redshift bin lies also $\sim2\sigma$ above the prediction for the high EBL model (KN high). Higher statistics and a closer look at the individual sources, which is beyond the scope of this paper, will be needed to investigate this further.

\section{Summary \& Conclusion}\label{summary}

In this paper the HE data from the 1FGL catalog is investigated for an attenuation signature resulting from the interaction of HE $\gamma$-rays and low energy photons from the EBL. Such an EBL attenuation signature is expected to show up in the energy range 10-100\,GeV and should exhibit a characteristic redshift dependence. The catalog provides a homogenous sample of HE data which, for the first time, enables a population study on the EBL attenuation effect on such a sample. Two different methods have been applied to search for such a signature but no significant signal has been detected. The expected signal is rather small and the current statistic in the catalog is not sufficient for a strong detection with the methods applied.

The 1FGL contains results from 11 month of data taking. Fermi is expected to operated for five to ten years which will significantly increase the statistics. The error on the power law fit and on the flux in the last bin $F_{10-100\,GeV}$ will improve by a factor $\sqrt{5}\sim 2$ to $\sqrt{10} \sim 3$. The increased statistic will also lead to the significant detection in the highest energy bin for many of those sources for which at the moment only an upper limit is available. The number of sources having a significant detection in the highest energy bin after 5/10 years of observations can be estimated using the results from the power law fit given in the catalog:  adopting a minimum flux of $F_{10-100\,GeV} = 1.5 \cdot 10^{-10}$\,cm$^{-2}$\,s$^{-1}$ for a significant detection in the 11 month data-set\footnote{This approximately corresponds to the lowest flux with a significant detection in the last bin for the 11 months data-set.} it can be estimated that after 5 (10) years of observations $\sim$50\% ($\sim$65\%) of the AGN sources in the 1FGL will have a significant detection at 10-100\,GeV compared to 20\% after 11 months.  The increased observation time will also lead to the discovery of new sources not yet included in the 1FGL. Of the $\sim$700 sources in the 1FGL associated with extragalactic sources only $\sim$55\% have a known redshift. Dedicated multi-wavelength observation campaigns are being performed to improve the data available for the associated sources (e.g. redshift). In addition, the understanding of the systematics at the high energy end of the Fermi/LAT energy regime, which, for the moment, has possibly still large uncertainties, will greatly improve.

Additional information can be derived by combining the Fermi/LAT HE data with data from lower and higher energies. Simultaneous observations at low energies (radio/optical/x-lrays) enable to constrain the modeling of the sources and thereby enables to derive constraints on the high energy part of the spectrum \citep{mankuzhiyil:2010a}. Combining the HE data with VHE observations from ground-based instruments gives a handle on the spectrum with and without EBL attenuation \citep{abdo:2009:fermi:tevselectedagn}, although this type of study is limited by the number of sources detected with both instruments and the difference in sensitivity. This situation will improve with the upcoming generation of ground-based VHE instruments (MAGIC II, H.E.S.S. II) which focus on the energy regime between 10 and 100\,GeV.

\medskip

\noindent \textit{Note:} After submission of this article (May 3, 2010) an article by the Fermi collaboration appeared on the preprint server \citep{abdo:2010:fermi:eblconstrains} which discusses, among other methods, a similar investigation to search for an EBL attenuation signature in the Fermi catalog data, which gives comparable results.


\begin{acknowledgements}
The author would like to thank D. Horns for fruitful discussions and D. Mazin for helpful comments.  The author would also like to thank the referee for helpful suggestions and comments.
\end{acknowledgements}


\def\Journal#1#2#3#4{{#4}, {#1}, {#2}, #3}
\def\NAT{Nature}
\def\AAA{A\&A}
\def\ApJ{ApJ}
\def\AJ{Astronom. Journal}
\def\Aph{Astropart. Phys.}
\def\ApJS{ApJSS}
\def\MNRAS{MNRAS}
\def\NIM{Nucl. Instrum. Methods}
\def\NIMA{Nucl. Instrum. Methods A}


\bibliographystyle{aa}

\end{document}